\documentclass[aps,pra,twocolumn,showpacs,groupedaddress,amsmath,amssymb]{revtex4}

\usepackage{graphicx}

\usepackage{color}
\usepackage{comment}
\usepackage{bm}
\usepackage{latexsym}

\begin{document}

\title{Observation of Asymmetric Transport in Structures with Active Nonlinearities}
\author{N. Bender, S. Factor, J. D. Bodyfelt, H. Ramezani, F. M. Ellis, T. Kottos} 
\affiliation{Department of Physics, Wesleyan University, Middletown, CT-06459, USA}
\date{\today}

\begin{abstract}
A mechanism for asymmetric transport based on the interplay between the fundamental symmetries of parity 
(${\cal P}$) and time (${\cal T}$) with nonlinearity is presented. We experimentally demonstrate and
theoretically analyze the phenomenon using a pair of coupled van der Pol oscillators, as a reference system,  
one with anharmonic gain and the other with complementary anharmonic loss; connected to two transmission 
lines. An increase of the gain/loss strength or the number of ${\cal PT}$-symmetric nonlinear 
dimers in a chain, can increase both the asymmetry and transmittance intensities. 
\end{abstract}

\maketitle


Directed transport is at the heart of many fundamental problems in physics. Furthermore it 
is of great relevance to engineering where the challenge is to design on-chip integrated devices that control energy 
and/or mass flows in different spatial directions. Along these lines, the creation of novel classes of integrated photonic, 
electronic, acoustic or thermal diodes is of great interest.  Unidirectional elements constitute the basic building blocks 
for a variety of transport-based devices such as rectifiers, pumps, molecular switches and transistors. 

The idea was originally implemented in the electronics framework, with the construction of electrical diodes that 
were able to rectify the current flux. This significant revolution motivated researchers to investigate 
the possibility of implementing this idea of "diode action" to other areas. For example, a proposal for the creation of a 
thermal diode, capable of transmitting heat asymmetrically between two temperature sources, was suggested in Ref. 
\cite{TPC02}. Another domain of application was the propagation of acoustic pulses in granular systems \cite{NDHJ05}.

A related issue concerns the possibility of devising an optical diode which transmits light differently along opposite propagation 
directions. Currently, such unidirectional elements rely almost exclusively on the Faraday effect, where external magnetic fields 
are used to break space-time symmetry. Generally this requires materials with appreciable Verdet constants and/or large size 
non-reciprocal devices -- typically  not compatible with on-chip integration schemes or light-emitting wafers \cite{ST91}. To 
address these problems, alternative proposals for the creation of optical diodes have been suggested recently. Examples include 
optical diodes based on second harmonic generation in asymmetric waveguides \cite{GAPF01} and nonlinear 
photonic crystals \cite{SDBB94}, photonic quasi-crystals and molecules \cite{B08}, or asymmetric nonlinear structures \cite{LC11}. 
Most of these schemes, however, suffer from serious drawbacks making them unsuitable for commercial or small-scale applications. 
Relatively large physical sizes are often needed while absorption or direct reflection dramatically affects the functionality leading to 
an inadequate balance between figures of merit and optical intensities. In other cases, cumbersome structural designs are 
necessary to provide structural asymmetry, or the transmitted signal has different characteristics than the incident one.  

In this Letter we demonstrate, experimentally and theoretically, a mechanism for asymmetric transport exploiting the co-existence 
of active elements with distinctive features of nonlinear dynamical systems, such as amplitude-dependent resonances. As a 
reference model we will use coupled nonlinear electronic Van der Pol (VDP) oscillators with anharmonic parts consisting of complementary 
amplifiers (gain) and dissipative conductors (loss) combined to preserve parity-time (${\cal PT}$) symmetry. ${\cal PT}$-symmetric 
structures were inspired by quantum field theories \cite{BB98}; their technological importance was first recognized in the 
framework of optics \cite{MGCM08}, where several intriguing features were found \cite{MGCM08,GSDMVASC09,RMGCSK10,
K10,RKGC10,ZCFK10,L09,L10b,M09,LRKCC11,BFKS09,HRTKVKDC12}. For example, the theoretical proposal of Refs.~\cite{RKGC10,
LRKCC11} suggested using nonlinearities to induce asymmetric transport. Very recently the idea of creating ${\cal PT}$-symmetric 
devices within the electronics framework was proposed and experimentally demonstrated in Ref.~\cite{SLZEK11}.  ${\cal PT}$-electronics 
provides a platform for detailed scrutiny of many new concepts within a framework of easily accessible experimental configurations 
\cite{SLZEK11,LSEK12,RSEGK12}. Despite all this activity, the ${\cal PT}$-symmetric Hamiltonians introduced in  quantum field theory, optics, 
and electronics have been restricted to conservative anharmonic constituents (if any) with the matched gain and loss 
exclusively linear (see however the theoretical works \cite{MMK11}).

\begin{figure}
\includegraphics[width=1\linewidth, angle=0]{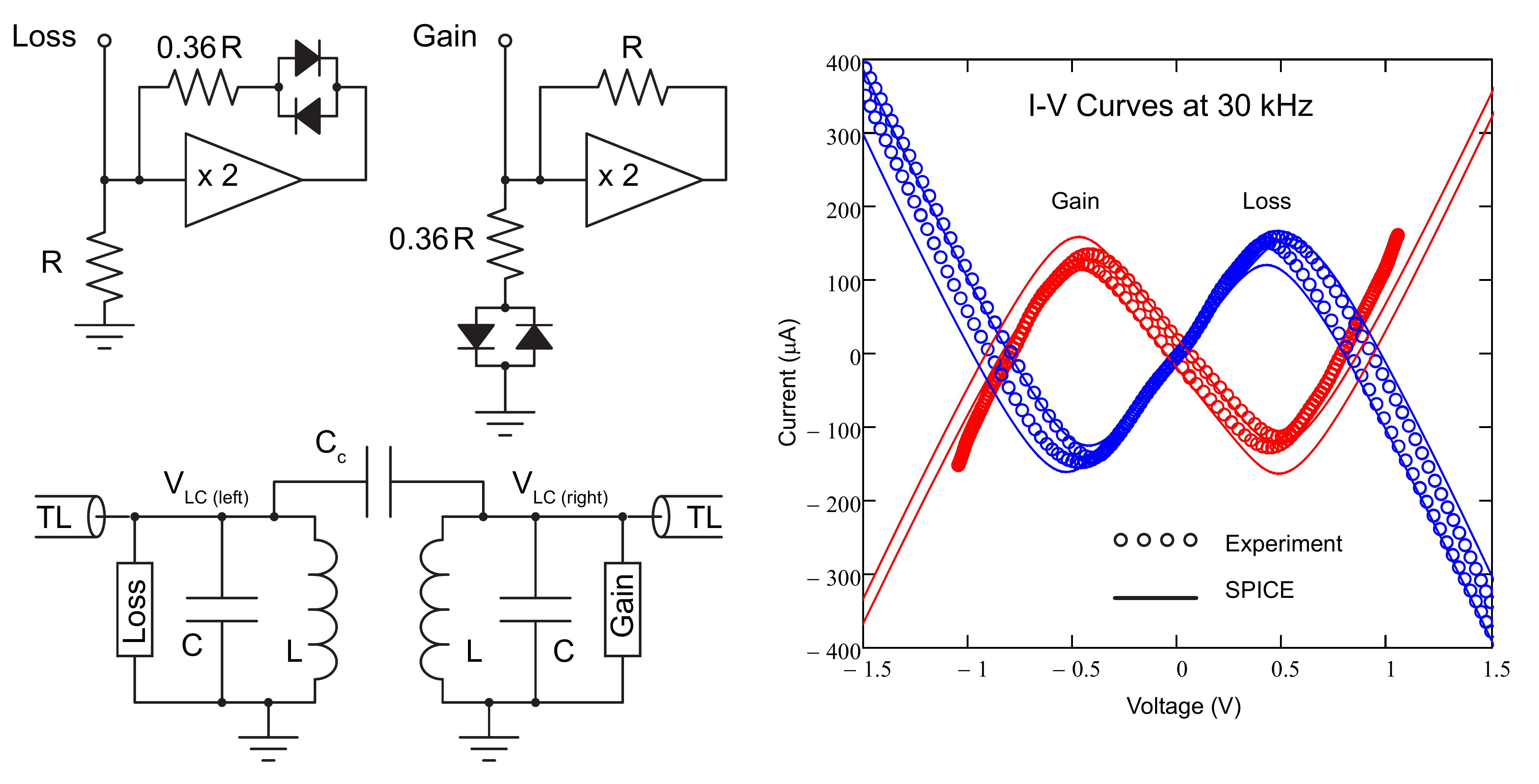}
\caption{(Left) Gain and loss circuits of the van der Pol ${\cal PT}$-symmetric dimer. The elements shown above are incorporated 
as parallel conductances in the capacitively coupled $LC$ resonators making up the dimer. (Right, color 
online) Experimental I-V response (circles) for the gain (red) and loss (blue) elements along with the 
corresponding NGSPICE simulations (solid), taken at a frequency of $30kHz$, typical of the active range of the VDP dimer.
}
\label{fig:fig1}
\end{figure}

Here, using the ${\cal PT}$-electronics framework, we demonstrate with experiment, simulations, and theory, asymmetric 
transport from ${\cal PT}$-symmetric structures that belong to this relatively unexplored class of nonlinear systems whose 
anharmonic parts includes the mutually matched gain and loss. Such systems, natural to the realm of electronics for which 
the van der Pol model was originally proposed \cite{vdp27}, can also be implemented in optics using, for example, optical amplifiers,
saturable absorbers \cite{K96} and two-photon losses to realize the nonlinear ${\cal PT}$-symmetry.

An ideal VDP oscillator has a linear anti-damping at low amplitudes which is subsequently overtaken by a cubic dissipation
at high amplitudes. In electronics this is an LC oscillator in parallel with a voltage dependent conductance 
characterized by the I-V curve $I(V)=-V/R+bV^3$. A negative impedance converter (NIC) (upper left of Fig. ~\ref{fig:fig1}) 
generates a $-1/R$ term, and we approximate the cubic turn-around with parallel back-to-back diodes
moderated by a resistor. The time-reversed conductance is constructed with the resistor $R$ and the diode combination 
interchanged. The resulting "gain" and "loss" nonlinear conductances refer to their low amplitude character. The 
respective nonlinear I-V curves are shown in the right panel of Fig.~\ref{fig:fig1}. The $0.36R$ in series with the diodes 
optimizes the match to a cubic nonlinearity. It is important to note that only the parameter $R$ is used to set the 
gain/loss parameter $\gamma = R^{-1}\sqrt{L/C}$, while the diode turn-on characteristics are fixed. When comparisons 
are made to theoretical models, the voltage scaling will consequently depend on $\gamma$.

The schematic of the dimer circuit is shown in the lower left of Fig.~\ref{fig:fig1}. The coupled $LC$ heart of the circuit is identical to 
that used in a previous work \cite{SLZEK11} with LM356 op-amps serving in the NICs and signal buffers. The two VDP oscillators 
are capacitively coupled by $C_c$. In principle, more complicated geometries such as chains of active nonlinear ${\cal PT}$-
symmetric dimers could be constructed by capacitively coupling additional dimers into the desired topology. The experimental 
measurements of this work are limited to the dimer as a check on SPICE simulations used for larger systems.

Transmission lines (TL) with impedance $Z_0$ are attached to the left (lossy) and the right (gain) $LC$ nodes of the dimer 
to complete the scattering system used to perform our transport measurements. Experimentally, these take the form of resistances 
$R_0=Z_0$ in series with independent voltage sources, here HP3325A synthesizers, on the right and left sides. The incoming 
and outgoing traveling wave components associated with a particular TL are deduced from the complex voltages on both ends 
of $R_0$, as sampled by a Tektronix DPO2014 oscilloscope. For example, on the left (lossy) side, with $V_{LC}$ the voltage 
amplitude on the $LC$ circuit node, and $V_0$ the voltage amplitude on the synthesizer side of the coupling resistor $R_0$, 
the incident wave on the dimer has a voltage amplitude $V_L^{+}=V_0/2$ and the outgoing wave has a voltage amplitude $V_L^{-}
=V_{LC}-V_0/2$. An equivalent relation holds for the right TL terminal with the $\pm$ superscripts interchanged, since they  
refer to right or left wave traveling direction regardless of the terminal orientation. 

The scattering measurements are performed for fixed incoming wave amplitude set by $V_0$ of the signal generator on either 
the left or the right side with the other side set to zero. The generator frequency is stepped (up or down), and the three relevant 
waveforms, $V_0$ and $V_{LC}$ on the left and right are simultaneously captured (the $V_0$ channel on the transmitted side 
is zero). Harmonic components of each wave constituent can be independently analyzed for magnitude, and relative phase. 
Instrumentation noise and sample time determine the accuracy of this analysis, which was found to be $<1\%$.

Circuit behavior was numerically modeled by the NGSPICE simulator \cite{ngspice}.  Modern simulation goes beyond a lumped element 
approach, in which circuits can only be modeled by idealized passive component approximations. Rather, behavioral modeling focuses 
on the mathematical relations between inputs, parameters, and outputs of a complicated nonlinear component; often by incorporating 
virtual dependent sources into the resulting model. This allows for faster speed and higher accuracy in calculations of circuit 
behavior. Most manufacturers, in fact, provide ready-made behavioral models for their devices. Here, for the LF356 op-amp we heavily 
utilized Texas Instruments' modified Boyle model, as well as the parametrized diode model of Fairchild's 1N914 \cite{model_src}. 

In Fig.\ref{fig:fig1}, circuit analysis was done in the time-domain for each individual element (gain or loss circuit) of the nonlinear 
${\cal PT}$-symmetric dimer. Using initial DC operating conditions, a frequency-dependent source drives a circuit through a transient 
regime into steady-state operation, at which point the voltages and currents are recorded. The measured I-V response curve (see
Fig. \ref{fig:fig1}) is reproduced by the simulations.

\begin{figure}
   \includegraphics[width=.7\linewidth, angle=0]{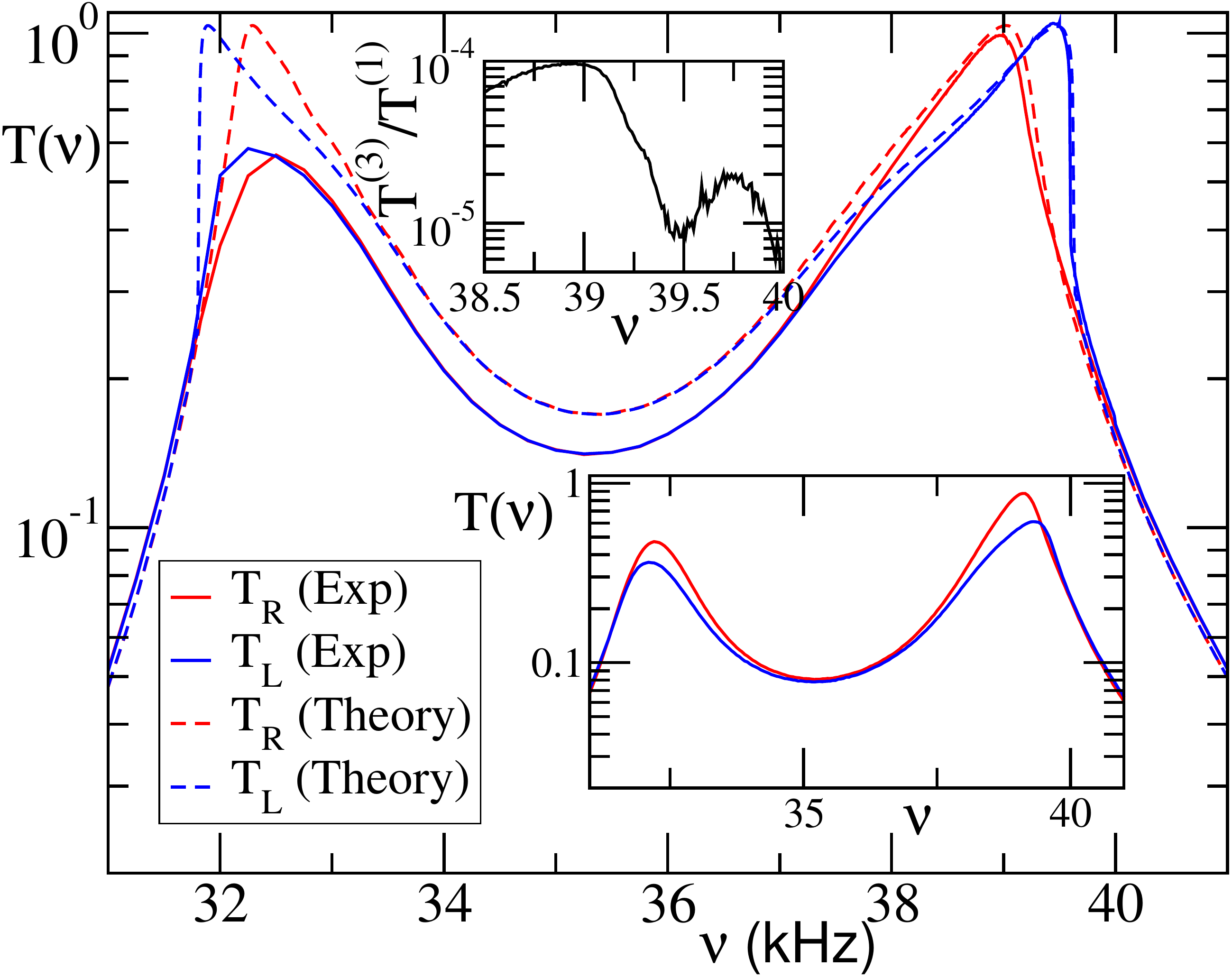}
    \caption{(Color online) Experimental data of transmittances $T_L$ and $T_R$ for a left (solid blue line) and right 
(solid red line) incident waves. The results of the numerical simulations with NGSPICE for $T_{L} (T_R)$ transmittances 
are shown as dashed blue (red) lines. Upper inset: The ratio between the transmittance associated with the third 
harmonic $T^{(3)}$ and the fundamental frequency $T^{(1)}$. Here, $\eta = 0.031$ and $\gamma = 0.15$. Lower inset: 
Experimental $T(\nu)$ for the same $\eta$ but smaller gain/loss parameter $\gamma = 0.11$. Note that as $\gamma$ decreases, 
the asymmetry and the transmitted intensity are reduced.}
\label{fig:fig2}
\end{figure}

For the transmittances $T(\nu)$ versus the driving frequency $\nu$ in Fig.\ref{fig:fig2}, the steady-state time-domain simulations of 
the ${\cal PT}$-symmetric dimer in the scattering configuration are similarly obtained. Fourier analysis is used to extract the relevant 
frequency-dependent voltages and currents, which are then used to calculate the scattering parameters. As a check of the accuracy 
of our numerical approach, we have also extracted the transmittance  using a nonlinear harmonic balance circuit analysis 
\cite{hb_footnote}. We have confirmed that the results are identical, within numerical accuracy, to the ones obtained from the 
time-domain analysis. In Fig. \ref{fig:fig2} we also report the experimental left and right transmittances for the 
${\cal PT}$-symmetric VDP dimer. The overall shape of the measured transmittances reasonably match the numerical simulations. The 
deviations are likely associated with a small parasitic inductive coupling.

A striking feature of the results of Fig. \ref{fig:fig2} is the fact that the transmittance from left to right $T_L(\nu)$ differs 
from the transmittance from right to left $T_R(\nu)$, i.e. $T_L\neq T_R$. The phenomenon is more pronounced closer to 
the resonance frequencies of the corresponding linear problem and is the main result of this paper. This asymmetry is 
forbidden by the reciprocity theorem in the case of linear, time-reversal symmetric systems.  In fact, it 
is not present even in the case of linear ${\cal PT}$-symmetric structures \cite{LSEK12}, although the time-reversal 
symmetry in this case is also broken. At the same time, a Hamiltonian nonlinear medium by itself cannot generate such transport 
asymmetries. Furthermore, we find that increasing the gain/loss parameter $\gamma$ which is responsible for the 
asymmetric transport, maintains or even enhances the transmitted intensities (compare the lower inset of Fig. \ref{fig:fig2}
with the main panel). This has to be contrasted with other proposals of asymmetric transport (see for example Ref. \cite{LC11})
where increase of asymmetry always leads to reduced transmittances.

We have also developed a theoretical understanding of asymmetric transport by restricting our analysis to the basic harmonic. This 
approximation is justified by our measurements which indicate that the main part of the transmitted 
power is concentrated in the fundamental harmonic. An example is shown in the upper inset of Fig. \ref{fig:fig2} 
where we report the ratio $T^{(3)}/T^{(1)}$ between transmittances of the third harmonic to the fundamental. 
Even harmonics are absent in the transmission spectra due to the nature of VDP anharmonicity, while for higher harmonics 
$T^{(n>3)}$ the experimental values of $T^{(n)}/T^{(1)}$ are  below the noise level of our measurements. 

Application of the first and second Kirchoff's laws at the TL-dimer contacts allow us to find the current/voltage wave 
amplitudes $I,V$ at the left (L) and right (R) contact. We get 
\begin{eqnarray}
\eta \frac{d\mathcal{I}_{L}}{d \tau} = \gamma(1 - \mathcal{V}_{L}^{2})\frac{d\mathcal{V}_{L} }{d \tau} + \mathcal{V}_{L} + 
( 1  + c)\frac{d^{2}\mathcal{V}_{L} }{d \tau^{2}}  - c \frac{d^{2}\mathcal{V}_{R} }{d \tau^{2}},\label{Kirchoff}\\
\eta \frac{d\mathcal{I}_{R}}{d \tau} = \gamma (1 - \mathcal{V}_{R}^{2})\frac{d\mathcal{V}_{R} }{d \tau} - \mathcal{V}_{R} - 
( 1  + c)\frac{d^{2}\mathcal{V}_{R} }{d \tau^{2}}  + c \frac{d^{2}\mathcal{V}_{L} }{d \tau^{2}}\nonumber
\end{eqnarray}
where the dimensionless current/voltage amplitudes ${\cal I},{\cal V}$ at the lead-dimer contacts are defined as $I_{L/R} = 
\frac{V_{0}}{Z_{0}} \mathcal{I}_{L/R}$, $V = V_{0} \mathcal{V}_{L/R}$. The dimensionless time is $\tau = t/\sqrt{LC}$ 
and $\eta=\sqrt{L/C}/Z_0$ is the dimensionless TL conductance, while we have also introduced the dimensionless capacitance $c=C_c/C$, 
a measure of the intra-dimer coupling.

At any point along a TL, the current and voltage determine the amplitudes of the right and left traveling wave components. The 
forward ${\cal V}_{L/R}^{+}$ and backward ${\cal V}_{L/R}^{-}$ wave amplitudes, and the voltage ${\cal V}_{L/R}$  and current 
${\cal I}_{L/R}$ at the TL-dimer contacts satisfy the continuity relation
\begin{eqnarray}
\label{scatstates}
{\cal V}_{L/R}&=&\left( {\cal V}_{L/R}^{+}+ {\cal V}_{L/R}^{-}\right) e^{-i\omega \tau}+cc;\\
{\cal I}_{L/R}&=&\left({\cal V}_{L/R}^{+}-{\cal V}_{L/R}^{-}\right) e^{-i\omega \tau}+cc\nonumber.
\end{eqnarray}

Note that Eqs. (\ref{Kirchoff}) contains nonlinear terms on the r.h.s. which are responsible for harmonic
generation. However, as suggested by the experimental data, we can neglect these higher harmonics and thus restrict our analytic study to 
the fundamental. Keeping this in mind when substituting Eqs. (\ref{scatstates}) into Eqs. (\ref{Kirchoff}), 
we get
\begin{widetext}
\begin{eqnarray}
(-1)^{s}i\omega\eta (-\mathcal{V}^{+}_{L/R} + \mathcal{V}^{-}_{L/R}) &=& \left[1-\omega^2(1+c)+i(-1)^{s+1}\omega\gamma 
(1-|\mathcal{V}^{+}_{L/R} + \mathcal{V}^{-}_{L/R}|^{2}) \right] (\mathcal{V}^{+}_{L/R} + \mathcal{V}^{-}_{L/R}) 
+ c\omega^{2} \mathcal{V}^{+/-}_{R/L}\nonumber\\
i\omega\eta(\mathcal{V}^{+/-}_{R/L}) &=& \left[(-1)^{s}i\omega\gamma (1 - |\mathcal{V}^{+/-}_{R/L}|^{2})-(1+c)\omega^{2}+1\right] 
\mathcal{V}^{+/-}_{R/L} +c\omega^2 (\mathcal{V}^{+}_{L/R} + \mathcal{V}^{-}_{L/R})
\label{bm}
\end{eqnarray}
\end{widetext}
where we used the compact notation $V_{R/L}^{+/-}$ for $V_{R}^{+}$ and $V_{L}^{-}$. The exponent $s$ above takes the values
$s=0,1$ for $L,R$ current amplitudes respectively.

\begin{figure}
\includegraphics[width=0.8 \linewidth, angle=0]{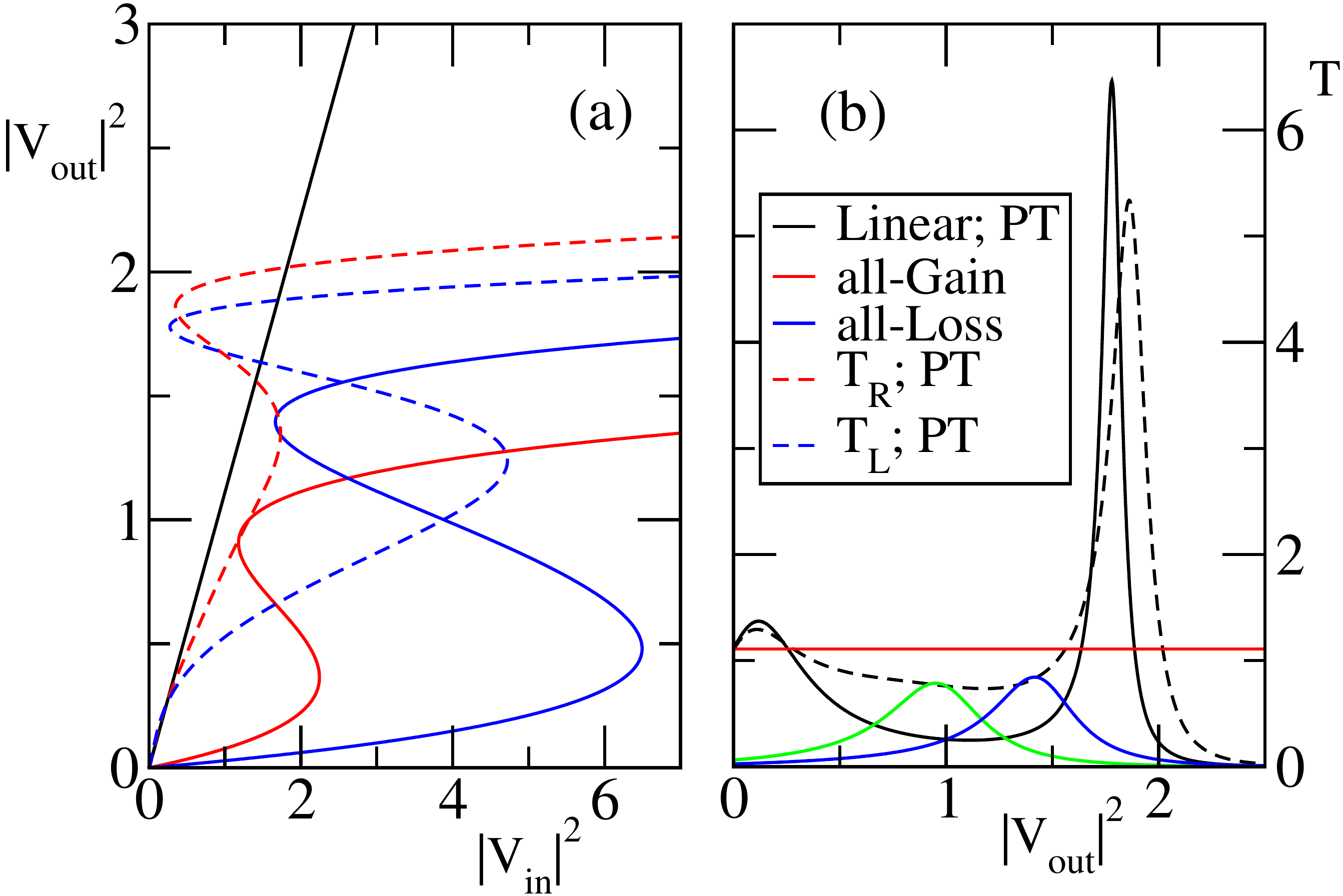}
\caption{(Color online) (a) Transmission curves for a nonlinear ${\cal PT}$-symmetric VDP dimer with $\gamma=0.151$, 
$\eta=0.0305$, $\omega=0.81$, and $c=0.267$. The linear ${\cal PT}$-symmetric case (black line) is reported also 
for comparison together with the nonlinear all-gain (red solid line) and all-loss (blue solid line) VDP dimer 
where reciprocal transport $T_L=T_R$ is found; (b) Corresponding transmittances for the cases of subfigure (a).
}
\label{fig:fig3}
\end{figure}

The way to solve Eqs.~(\ref{bm}) is to introduce the backward transfer map \cite{TH99}. The latter uses the output amplitude 
${\cal V}_{\rm out}={\cal V}_R^{+} ({\cal V}_L^{-})$ as an initial 
condition together with the boundary conditions 
${\cal V}_R^{-}=0$ (${\cal V}_L^{+}=0$) for a left (right) incoming wave. Iterating backwards, we calculate the corresponding 
incident ${\cal V}_{\rm in}={\cal V}_L^{+}$ (${\cal V}_R^{-}$) and reflected ${\cal V}_{\rm refl}={\cal V}_L^{-}$ 
(${\cal V}_R^{+}$) amplitudes for a left (right) incident wave. Representative $(|{\cal V}_{\rm in}|^2,|{\cal V}_{\rm out}|^2)$ 
curves are shown in Fig. \ref{fig:fig3}a. The associated transmittances are defined as $T=\left|{\cal V}_{\rm out}/
{\cal V}_{\rm in}\right|^2$ . Straightforward algebra gives:
\begin{equation}
T_{L} = \Big|\dfrac{2 \omega \eta c}{ \eta \alpha + \alpha\big(\gamma(1 -|\frac{\alpha{\cal V}_{\rm out}}{c\omega}|^2) + \frac{i}{\omega} - 
i \omega(1 + c)\big) + (c \omega)^2}\Big|^{2},\quad\label{TL}
\end{equation}
where $\alpha = \big(\eta - \gamma(1- |\mathcal{V}_{out}|^{2}) + \frac{i}{\omega} - i\omega(1+c)\big)$. The transmittance $T_{R}$ 
for a right incident wave is given by the same expression as Eq. (\ref{TL}) with the substitution of $\gamma\rightarrow -\gamma$ 
i.e. $T_{R}(\gamma)=T_{L}(-\gamma)\neq T_{L}(\gamma)$. The origin of the latter inequality is due to the fact that nonlinear 
resonances are detuned differently for left and right incident waves, as seen from Eq. (\ref{TL}) (compare black dashed and solid 
lines in Fig. \ref{fig:fig3}b). The case $\gamma=0$ corresponds to a linear passive dimer for which $T_L=T_R$ in agreement with 
Eq. (\ref{TL}). To further highlight the importance of the interplay between nonlinearity and ${\cal PT}$-symmetry, we also report 
in Fig. \ref{fig:fig3} the transmittances for a nonlinear VDP dimer with both elements having gain or loss and for a linear 
${\cal PT}$-symmetric dimer. In contrast to the nonlinear ${\cal PT}$-symmetric structure, the transmission is symmetric.

\begin{figure}[h!]
   \includegraphics[width=.85\linewidth, angle=0]{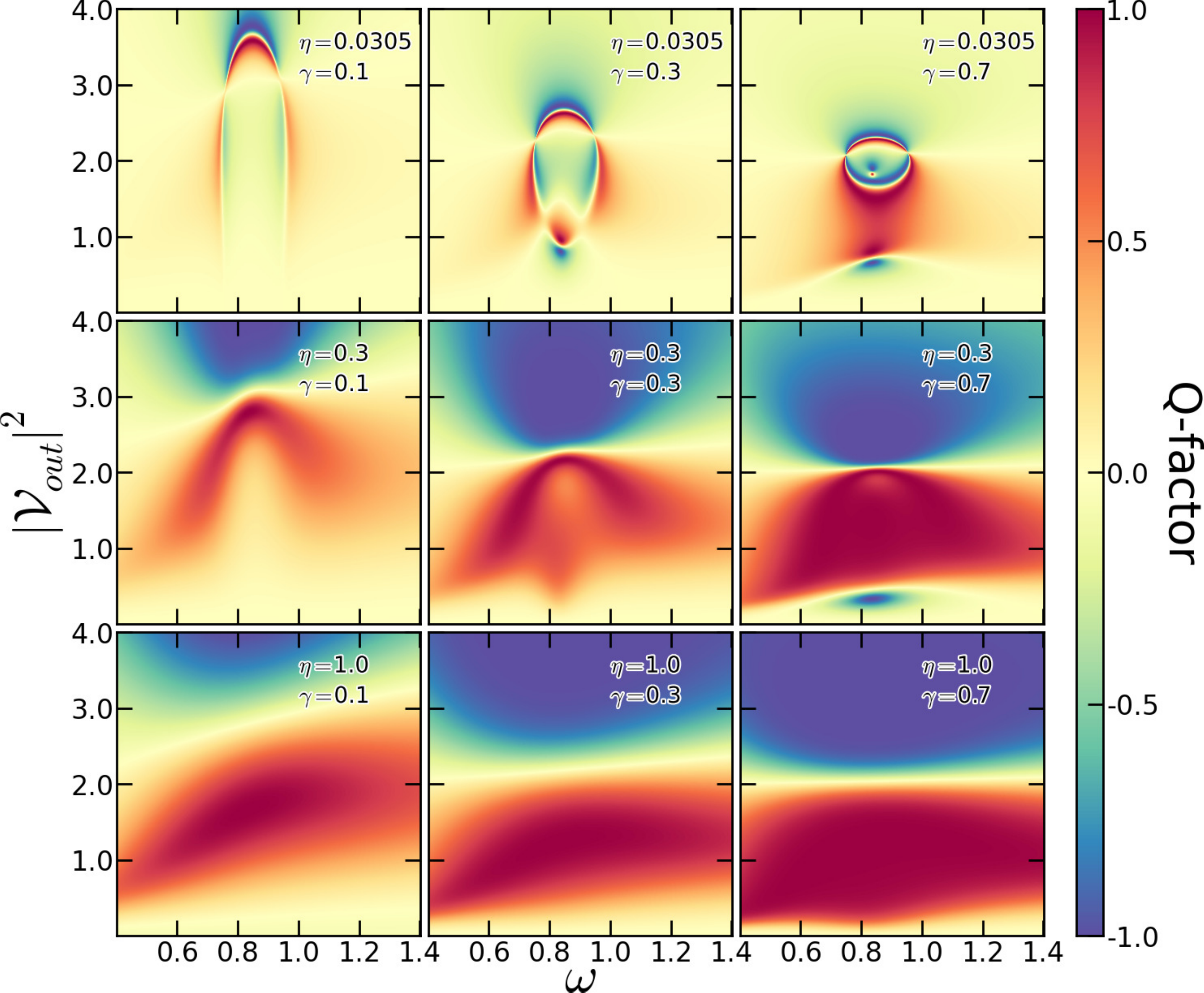}
    \caption{(Color online) Density plots of the $Q$-factor Eq. (\ref{Qfactor}) for increasing $\gamma$ (from left to right)
and $\eta$ (top to bottom). The $\eta,\gamma$ values are indicated in the subfigures while $c=0.267$.
\label{fig:fig4}}
\end{figure}

\begin{figure}
   \includegraphics[width=.75\linewidth, angle=0]{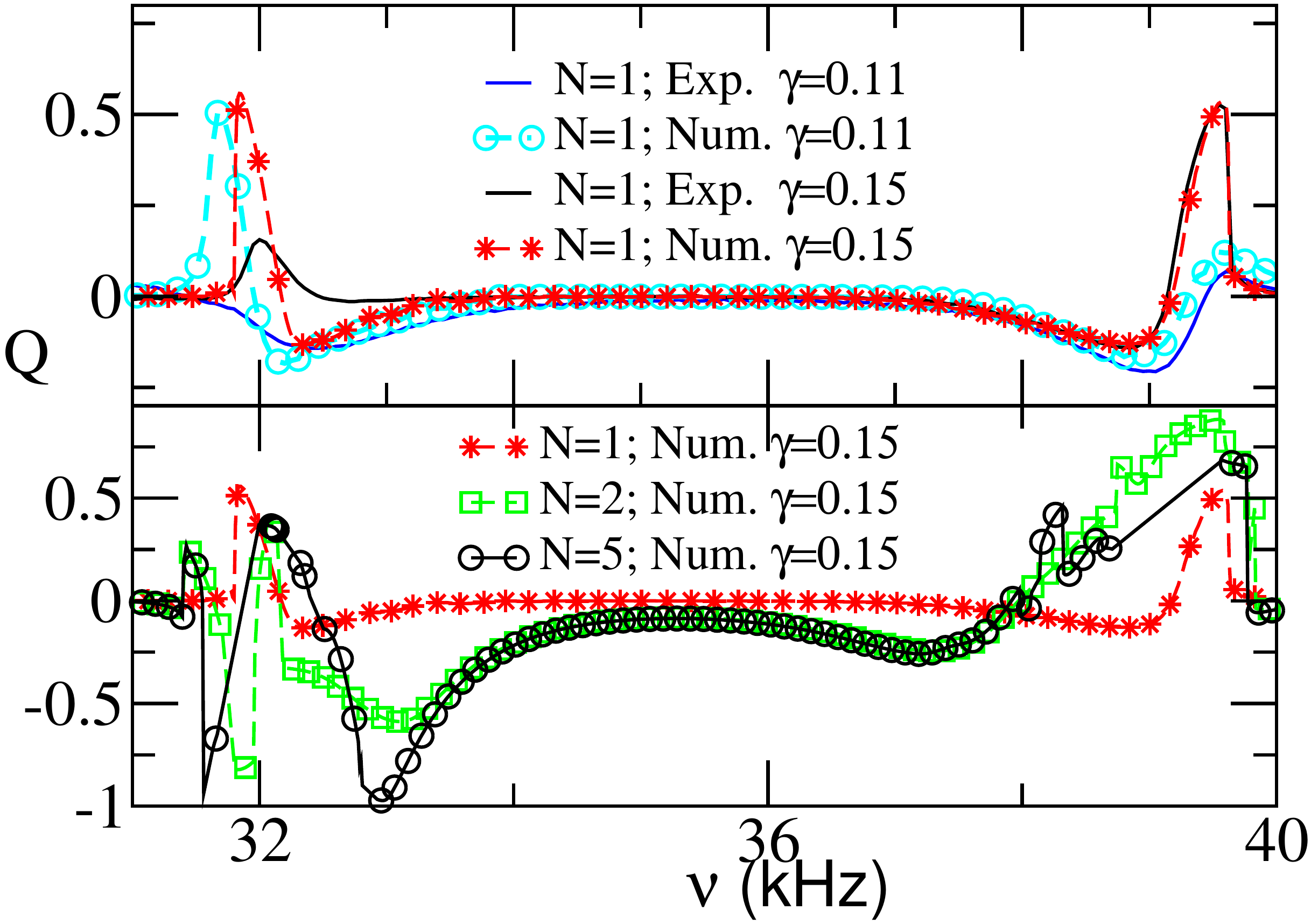}
    \caption{(Color online) (Upper) Experimental (solid lines) and SPICE (symbols) results for the rectification factor $Q$ Eq. (\ref{Qfactor}) for one 
($N=1$) and two ($N=2$) coupled ${\cal PT}$-symmetric VDP dimers for two different $\gamma$ values and $\eta=0.03$.  (Lower)
The SPICE simulations for $Q$ for moderate chains of VDP dimers with $N=1,2,5$ (we note that for $\nu\approx 38.5 kHz$ the simulations
were not converging for $N=5$). As $N$ or $\gamma$ increase the $Q$-factor increases as well.}
\label{fig:fig5}
\end{figure}

Since the phenomenon is nonlinear the asymmetry depends on both frequency and amplitude. To quantify its efficiency,
in Fig. \ref{fig:fig4} we report the rectification factor
\begin{equation}
\label{Qfactor}
Q=\frac{T_L({\cal V}_{\rm out})-T_R({\cal V}_{\rm out})}
{T_L({\cal V}_{\rm out})+T_R({\cal V}_{\rm out})}
\end{equation}
which approaches $\pm 1$ for maximal asymmetry. Note that increasing $\gamma,\eta$ broadens the regions in which $|Q|$ is
relatively large. 
Some representative experimental rectification factors $Q$ for a single dimer $N=1$ and two different values of $\gamma$ are
shown in Fig. \ref{fig:fig5} together with the SPICE simulations. The measurements and the simulations compare nicely with one 
another and are in agreement with the theoretical calculations shown in Fig. \ref{fig:fig4}. Furthermore, increasing 
the number $N$ of nonlinear ${\cal PT}$-symmetric dimers leads to a considerable increase of the $Q$-factor. This can be seen 
in Fig. \ref{fig:fig5} even for the case of $N=2$.  Although this behavior can be formally analyzed by evaluating the $N^{th}$ 
order backward transfer map, a qualitative explanation can be provided based on our understanding of the $N=1$ dimer. Specifically, 
one expects that the increase of the number of ${\cal PT}$-symmetric dimers introduces $2N$ nonlinear resonances (which are 
detuned differently for left and right incoming waves), thus broadening the frequency domain for which asymmetric transport is 
present.

{\it Conclusions -} Using coupled ${\cal PT}$-symmetric VDP oscillators, we have demonstrated experimentally and theoretically 
a mechanism based on the co-existence of nonlinearity and ${\cal PT}$-symmetry which leads to asymmetric wave transport. Unlike 
the linear ${\cal PT}$ or passive nonlinear cases, this is achieved concurrent with a significant level of transmittance. These 
findings can be applied to photonic or phononic systems as a guide for optimizing design in order to achieve maximal asymmetry.

{\it Acknowledgments --} 
This research was supported by an AFOSR grant No. FA 9550-10-1-0433, and by an NSF ECCS-1128571 grant. NB\&SF acknowledge support
from Wesleyan Faculty/Student Internship grants.


\end{document}